\def\dfrac#1#2{{\displaystyle{#1\over#2}}}

\def\d{\mbox{d}}
\documentstyle[aps,multicol]{revtex}

\begin{document}
\draft

\title{Short-range particle correlations in dilute Bose gas}
\author{A. Yu. Cherny$^{*}$ and A. A. Shanenko$^{**}$}
\address{Bogoliubov Laboratory of Theoretical Physics,
Joint Institute for Nuclear Research, 141980, Dubna, Moscow region,
Russia}
\date{March 22, 2000}
\maketitle

\begin{abstract}
The thermodynamics of a homogeneous dilute Bose gas with an arbitrary strong
repulsion between particles is investigated on the basis of the exact
relation connecting the pair correlation function with the in-medium pair
wave functions and occupation numbers. It is shown that the
effective-interaction scheme which is reduced to the Bogoliubov model with
the effective pairwise potential, is not acceptable for investigating the
short-range particle correlations in a dilute strongly interacting Bose gas.
In contrast to this scheme, our model is thermodynamically consistent and
free of the ultraviolet divergences due to accurate treatment of the
short-range boson correlations. The equation for the in-medium scattering
amplitude is derived that makes it possible to find the in-medium
renormalization for the pair wave functions at short boson separations.
Low-density expansions for the main thermodynamic quantities are
reinvestigated on the basis of this equation. Besides, the expansions are
found for the interaction and kinetic energies per particle. It is
demonstrated that for the many-boson system of the hard spheres the
interaction energy is equal to zero for any boson density. The exact
relationship between the chemical potential and in-medium pair wave functions
is also established.
\end{abstract}

\begin{multicols}{2}

\section{Introduction and basic equations}
\label{1sec}

The well-known experiments with the magnetically trapped alkali
atoms~\cite{anddavbrad} have significantly renewed interest in the
theory of the Bose-Einstein condensation (see, e.g.,
Ref.~\cite{RMP}). In particular, it has recently been demonstrated by
the present authors~\cite{Ch1} that the customary way of
investigating a dilute Bose gas with a pure repulsive and arbitrary
strong interaction~\cite{Lee} is thermodynamically inconsistent. At
$n=N/V \to 0$ this way is known~\cite{Fet} to be reduced to the
Bogoliubov model~\cite{Bog1} with the ``bare" pairwise potential
$\Phi(r)$ replaced by an effective, ``dressed" one. This is why below
the approach of Ref.~\cite{Lee} is called the ``effective-interaction
method". The ``dressed" pairwise potential is usually derived by
summing the ladder diagrams and involves, as is assumed, all the
necessary information on the short-range spatial correlations of
bosons~\cite{Lee}. In the final expressions use of the effective
interaction results in substituting the exact scattering length $a$
for its Born approximation $a_0$~\cite{Fet}. This allows for
operating with strongly singular potentials, but at the price of loss
of the thermodynamic consistency. On the contrary, the
strong-coupling generalization of the Bogoliubov model proposed by
the present authors in the paper~\cite{Ch1}, is based on the {\it
variational} procedure and does not invoke any mean-field arguments.
Owing to this structure of the generalization we do not need to worry
about the thermodynamic consistency.

The trouble mentioned above gives rise to various misrepresentations
of the effective-interaction approach. For example, the condition of
the self-consistency leads to zero condensate depletion within the
pseudopotential model~\cite{Huang}. Another its manifestation is
an irrelevant picture of the pair boson correlation at short particle
separations. This important point calls for a comprehensive analysis
which has not been fulfilled in Ref.~\cite{Ch1} for reason of space.
Thus, in the present paper we continue reinvestigation of a dilute
Bose gas with an arbitrary strong repulsion between particles within
the model proposed in the paper~\cite{Ch1}, the short-range boson
correlations being of special interest now. The zero temperature is
under consideration below.

The formalism of the present paper is concerned with the reduced
density matrix of the second order (the 2-matrix) and its
eigenfunctions which we call, following Bogoliubov~\cite{Bog2}, the
in-medium pair wave functions. As the 2-matrix with its
eigenfunctions are not often discussed in the modern scientific
literature on the Bose-Einstein condensation, it is worth noting some
basic notations and formulas. The 2-matrix for the many-body system
of spinless bosons can be represented as~\cite{boglec}: $$
\rho_2({\bf r}_1^{\prime},{\bf r}_2^{\prime};{\bf r}_1,{\bf r}_2) =
{F_2({\bf r}_1,{\bf r}_2;{\bf r}_1^{\prime},{\bf r}_2^{\prime}) \over
N(N-1)}, $$ where the pair correlation function is given by
\begin{equation}
F_2({\bf r}_1,{\bf r}_2;{\bf r}_1^{\prime},{\bf r}_2^{\prime})=
\langle \hat{\psi}^{\dagger}({\bf r}_1) \hat{\psi}^{\dagger}({\bf r}_2)
    \hat{\psi} ({\bf r}_2^{\prime})\hat{\psi} ({\bf r}_1^{\prime})\rangle.
\label{2}
\end{equation}
Here $\hat{\psi} ({\bf r})$ and $\hat{\psi}^{\dagger}({\bf r})$
denote the boson field operators. Use of the pair correlation
function that differs from the 2-matrix only by the normalization
factor, is more convenient in the thermodynamic limit
($n=N/V=\mbox{const}$, $V\to\infty$) when $F_2({\bf r}_1,{\bf
r}_2;{\bf r}'_1,{\bf r}'_2)\propto 1$ while $\rho_2({\bf r}'_1,{\bf
r}'_2;{\bf r}_1,{\bf r}_2)\propto 1/V^{2}$. Recently it has been
found~\cite{Ch1,Ch2,Ch3} that for the uniform Bose gas, the system
with a small depletion of the zero-momentum state, the correlation
function (\ref{2}) can be written in the thermodynamic limit as
follows:
\begin{eqnarray}
F_2({\bf r}_1,{\bf r}_2;{\bf r}_1^{\prime},{\bf r}_2^{\prime})
&=&n_0^2\varphi^*(r)\varphi(r')\nonumber\\
&&+2n_0\int\frac{\d^3q}{(2\pi)^3}n_q
\varphi_{{\bf q}/2}^*({\bf r})\varphi_{{\bf q}/2}({\bf r}')
\nonumber\\
&&\phantom{+2n_0\int}\times\exp[i{\bf q}\cdot({\bf R}'-{\bf R})],
\label{3}
\end{eqnarray}
where ${\bf r}={\bf r}_1 - {\bf r}_2, \;{\bf R}=({\bf r}_1 +{\bf
r}_2)/2$ and similar relations take place for ${\bf r}^{\prime}$ and
${\bf R}^{\prime}$, respectively. In Eq.~(\ref{3}) $n_0=N_0/V$ is the
density of the particles in the zero-momentum one-boson state,
$n_q=\langle \hat{a}_{\bf q}^{\dagger} \hat{a}_{\bf q}\rangle$ stands
for the distribution of the uncondensed bosons over momenta,
$\varphi(r)$ and $\varphi_{{\bf q}/2}({\bf r})$ are pair wave
functions in medium in the centre-of-mass system.  Namely,
$\varphi(r)$ is the wave function of a pair of particles being both
condensed. In turn, $\varphi_{{\bf q}/2}({\bf r})$ denotes the wave
function of the relative motion of a pair of bosons with the total
momentum $\hbar{\bf q}$, this pair including one condensed and one
uncondensed particles.  So, Eq.~(\ref{3}) takes into account the
condensate-condensate and supracondensate-condensate pair states and
is related to the situation of a small depletion of the zero-momentum
one-boson state. For the pair wave functions we have
\begin{eqnarray}
\varphi(r)\!=\!1\!+\!\psi(r),\,\varphi_{{\bf p}}({\bf r})\!=\!
\sqrt{2}\cos({\bf p}\cdot{\bf r})\!+\!\psi_{{\bf p}}({\bf r})\,
(p\!\not=\!0),
\label{4}
\end{eqnarray}
where the scattering waves $\psi(r)$ and $\psi_{{\bf p}}({\bf r})$ obey
the boundary conditions for $r\to\infty$:
\begin{equation}
\psi(r)\to 0, \quad \psi_{{\bf p}}({\bf r}) \to 0
\label{4bound}
\end{equation}
which follow from the Bogoliubov principle of correlation
weakening~\cite{Bog2}.  The Fourier transforms of the functions $\psi(r)$ and
$\psi_{{\bf p}}({\bf r})$ can explicitly be expressed in terms of the Bose
operators $\hat{a}_{{\bf p}}^{\dagger}$ and $\hat{a}_{{\bf p}}$~\cite{Ch2}:
\begin{eqnarray}
{\psi}(k)=\langle \hat{a}_{{\bf k}}\,\hat{a}_{-{\bf k}}\rangle/n_0,
{\psi}_{\bf p}({\bf k})=\!\sqrt{\frac{V}{2 n_0}}
\frac{\langle \hat{a}^{\dagger}_{2{\bf p}}
\hat{a}_{{\bf p}+{\bf k}} \hat{a}_{{\bf p}-{\bf k}}\rangle}{n_{2p}}.
\label{25}
\end{eqnarray}
In the representation~(\ref{3}) the terms corresponding to the
supracondensate-supracondensate ``channel", are neglected, i.e. we
omit the contribution of pairs of the particles which are both
uncondensed. Besides, it is assumed that there are no bound states of
pairs of bosons, which is obviously realized for a purely repulsive
interaction between bosons.

The diagonal matrix element of the pair correlation function
(\ref{3}) is proportional to the pair distribution function
\begin{equation}
g(r)=F_2({\bf r}_1,{\bf r}_2;{\bf r}_1,{\bf r}_2)/n^2
\label{8}
\end{equation}
that can directly be observed in the scattering experiments.
Derivation of Eqs.~(\ref{3})-(\ref{25}) and detailed discussions can
be found in Ref.~\cite{Ch2}.

The two limiting cases $n\to0$ and $r\to0$ correspond to the
situation when the behaviour of two particles in medium is
determined by the ordinary two-body problem provided the pairwise
interaction $\Phi(r)$ is repulsive and goes to infinity
at short boson separations. In
particular, when $n\to0$ we have $(n-n_{0})/n\to0$ and, as it is
known since the Bogoliubov original paper and follows also from
Eqs.~(\ref{3}) and (\ref{8}),
\begin{equation}
g(r) \to [\varphi^{(0)}(r)]^2.
\label{5}
\end{equation}
Here $\varphi^{(0)}(r)$ is defined by $\varphi^{(0)}(r)=\lim_{n \to
0}\varphi(r)$ and obeys the ordinary two-body Schr\"odinger equation
in the centre-of-mass system~(\ref{twobody}) (see Ref.~\cite{Bog1}).
Similarly, $\varphi^{(0)}_{\bf p}({\bf r})=\lim_{n\to0}\varphi_{\bf
p}({\bf r})$ obeys the Schr\"odinger equation related to the
eigenvalue $\hbar^{2}p^{2}/m$ that corresponds to the relative motion
of two particles with the momentum $\hbar {\bf p}$. These conditions
should be satisfied in any theory that appropriately takes into
account the short-range correlations of particles. Below we show that
our model~\cite{Ch1} leads to the correct picture of the spatial
correlations in contrast to the effective-interaction approach that
leads to negative values of the pair distribution function (see
Sec.~\ref{effpot}).

Having at our disposal the distribution function $n_{k}$ and the set
of the pair wave functions $\varphi(r)$ and $\varphi_{{\bf p}} ({\bf
r})$, we are able to calculate the main thermodynamic quantities of
the system of interest. In particular, the mean energy per particle
is expressed in terms of $n_k$ and $g(r)$ via the well-known formula
(see, e.g. Ref.~\cite{boglec})
\begin{eqnarray}
\varepsilon=\int \frac{\d^3k}{(2\pi)^3} T_k \frac{n_k}{n}+
\frac{n}{2}\int\d^3r\,g(r)\Phi(r),
\label{7}
\end{eqnarray}
where $T_k=\hbar^2 k^2/2m$ is the one-particle kinetic energy, $m$ is
the ``bare" mass of particles, and $n=N/V$ stands for the boson
density.

The organization of this paper is as follows. In Sec.~\ref{1asec} we
give, for convenience, helpful information concerning the
classification of the pairwise potentials used in the ordinary
two-body problem. In Sec.~\ref{2sec} the Bogoliubov model of a weakly
interacting Bose gas is considered within a variational scheme. This
scheme yields the system of two equations connecting $n_k$ with
$\varphi(r)$. As to the supracondensate-condensate pair wave
functions, they are the symmetrized plane waves in the Bogoliubov
model: $\psi_{{\bf p}}({\bf r})=0.$ In the next section the
low-density expansions for the condensate depletion and mean energy
per particle of a weak-coupling Bose gas are calculated within the
Bogoliubov model. The effective-interaction approach of
Ref.~\cite{Lee} is analyzed in section~\ref{effpot}. Using the
results of the previous sections~\ref{2sec} and \ref{densbog}, we
show that the effective-interaction approach is thermodynamically
inconsistent. This inconsistency turned out to be directly related to
an irrelevant picture of the short-range spatial boson correlations.
In particular, for a strongly singular potential the
effective-interaction scheme yields for the pair distribution
function the nonphysical result $g(r=0)=-1$ in the limit $n\to 0$. It
is also demonstrated that the well-known ultraviolet divergence
appearing in the effective-interaction approach as well as the
thermodynamic inconsistency occur because the Bogoliubov framework is
used beyond the range of its validity. The regularizing procedure
which consists in omitting the divergent integral $\int
\d^{3}k/k^{2}$, can be justified provided the quantities of interest
depend on the pairwise potential through the mediation of the
scattering length (\ref{adef}). Sec.~\ref{genBog} regards a correct
strong-coupling generalization of the Bogoliubov model.  This
generalization is based on Eq.~(\ref{8}) taken together with
Eq.~(\ref{3}) rather than on its linearized variant Eq.~(\ref{11})
used in the effective-interaction approach and being the
weak-coupling approximation for $g(r)$.  A variational procedure
similar to that of Sec.~\ref{2sec} is formulated. It provides the
system of equations which should be solved to find the pair wave
functions in conjunction with the momentum distribution. For a dilute
Bose gas this system is reduced to the set of two equations
connecting $n_k$ and $\varphi(r)$. There is essential difference
between these equations and those of Sec.~\ref{2sec}. Now the
pairwise potential $\Phi(r)$ appears only in the combination
$\varphi(r)\Phi(r)$ which allows for using the strongly singular
potentials beyond the effective-interaction scheme. In what concerns
the supracondensate-condensate contribution to the thermodynamic
quantities, it can be calculated with the relation $\lim_{p \to
0}\varphi_{{\bf p}}({\bf r})= \sqrt{2} \varphi(r)$ resulting from
Eqs.~(\ref{4}) and (\ref{4bound}). In the next section we investigate
the short-range renormalization for $\varphi(r)$ conditioned by
presence of surrounding bosons. Its long-range behavior is also
discussed. All the investigation of this section is based on the
in-medium Lippmann-Schwinger equation coming from the equation for
$\varphi(r)$ found in the previous section. In Sec.~\ref{5sec} the
low-density expansions for the Bose condensate depletion, the energy
per particle and the chemical potential are found within the model
presented in Sec.~\ref{genBog}, various calculation ways being used.
For this purpose we establish the exact relationship between the
chemical potential and pair wave functions in a condensed many-boson
system.  Here we also evaluate the kinetic and interaction energies
per particle which, to our knowledge, have never been calculated. It
should be stressed that they explicitly depend on a shape of the
pairwise potential even in the leading order of the low-density
expansion. In the framework of our approach we are able to perform
all the calculations concerning the kinetic and potential energies
both directly and with the Hellmann-Feynman theorem, in contrast to
the effective-interaction method. Main results and prospects are
discussed in the last section.

\section{Classification of interaction potentials}
\label{1asec}

Before further consideration we recall the classification of the pairwise
interactions $\Phi(r)$ which is used in the ordinary two-body problem. In
this paper we only deal with the short-range potentials which go to zero
for $r\to \infty$ as $\Phi(r)\to 1/r^{m}$ ($m>3$), or even faster. Let us
consider the solution of the two-body Schr\"odinger equation in the
centre-of-mass system
\begin{equation}
-\frac{\hbar^2}{m}\nabla^2\varphi^{(0)}(r)+\Phi(r)\varphi^{(0)}(r)=0
\label{twobody}
\end{equation}
which corresponds to the scattering state with the momentum $p=0$:
$\varphi^{(0)}(r)=1+\psi^{(0)}(r)$, where the scattering part behaves as
\begin{equation}
\psi^{(0)}(r)\to -a/r
\label{scatasymp}
\end{equation}
when $r\to \infty$. Owing to this boundary condition with the real
quantity $a$, the solution $\varphi^{(0)}(r)$ is chosen to be real
also. The scattering length $a$ is defined by means of the scattering
amplitude $U^{(0)}(0)$:
\begin{eqnarray}
a
&=&\frac{m}{4\pi\hbar^{2}}U^{(0)}(0),
\label{adef} \\
U^{(0)}(0)
&=&\int\d^3r\,\varphi^{(0)}(r)\Phi(r).
\label{u0bare}
\end{eqnarray}
As applied to Eq. (\ref{twobody}), the perturbation technique gives the
expansion for its solution
\begin{eqnarray}
\psi^{(0)}(k)
&=&\psi^{(0)}_{1}(k)+\psi^{(0)}_{2}(k)+\cdots,
\label{exppsi}\\
\psi^{(0)}_{1}(k)
&=&-\Phi(k)/(2T_{k}),
\label{bornpsi}
\end{eqnarray}
which leads to the following expansion for the scattering length
(\ref{adef}):
\begin{eqnarray}
a
&=&a_{0}+a_{1}+a_{2}+\cdots,
\label{expa}\\
a_{0}
&=&\frac{m}{4\pi\hbar^{2}}\Phi(k=0),\;
a_{1}=-\frac{m}{4\pi\hbar^{2}}\int\frac{\d^{3}k}{(2\pi)^{3}}
\frac{\Phi^{2}(k)}{2T_{k}}.
\label{borna}
\end{eqnarray}
Here $\psi^{(0)}(k)$ and $\Phi(k)$ stand for the Fourier transforms
of $\psi^{(0)}(r)$ and $\Phi(r)$, respectively. If we restrict
ourselves to the first terms in Eqs.~(\ref{exppsi}) and (\ref{expa})
[$\psi^{(0)}(k)\simeq \psi^{(0)}_{1}(k)$ and $a\simeq a_{0}$] we
arrive at the Born approximation for the wave function and the
scattering length, respectively.

The interaction is called the weak-coupling one provided the Born
approximation works well, in particular,
\begin{equation}
\int\frac{\d^{3}k}{(2\pi)^{3}}\frac{\Phi^{2}(k)}{2T_{k}}
\ll \Phi(k=0).
\label{borncrit}
\end{equation}
This is valid if, first, the potential $\Phi(r)$ is integrable, and,
second, it is proportional to a small parameter, the coupling
constant. The latter implies that $|\psi^{(0)}(r)| \ll 1$, and, so,
the Born approximation (\ref{bornpsi}) is nothing else but
linearization of Eq.~(\ref{twobody}) with respect to $\psi^{(0)}(r)$
proportional to the coupling constant:
\[
\frac{\nabla^2\bigl(1+\psi^{(0)}(r)\bigr)}{\bigl(1+\psi^{(0)}(r)\bigr)}
\simeq \nabla^2\psi^{(0)}(r).
\]
The potential is called singular if it is integrable but the Born
approximation does not work well.  At last, the potential is of strongly
singular, or hard-core, type if it is not integrable [$\Phi(r)\to 1/r^{m}$
($m\geq 3$) for $r\to 0$], and the terms (\ref{bornpsi}) and (\ref{borna})
cannot thus exist.  In the present paper the pairwise interaction of this
type is exactly implied when we speak about the strong-coupling regime. For
example, the well-known Lennard-Jones potential corresponds to this case
together with the hard-sphere interaction
\begin{equation}
\Phi(r)=\left\{\begin{array}{ll}
+\infty, & r<a, \\
0,       & r>a.
\end{array}\right.
\label{hardsph}
\end{equation}

In the strong-coupling regime the solution of Eq.~(\ref{twobody})
obeys the boundary condition $\varphi^{(0)}(r=0)=0$, otherwise the
interaction energy $E_{int}=\int\d^{3} r\,[\varphi^{(0)}(r)]^{2}
\Phi(r)$ and the scattering length (\ref{adef}) would be infinitely
large.

In further consideration we make use of the variational theorem for the
scattering amplitude (\ref{u0bare}):
\begin{eqnarray}
\delta{U}^{(0)}(0)=\int\d^3r
\Bigl[
&&\psi^{(0)}(r)\delta\Bigl(-\frac{\hbar^2}{m}\nabla^2 \Bigr)\psi^{(0)}(r)
\nonumber \\
&&+\varphi^{(0)}(r)\delta\bigl(\Phi(r)\bigr)\varphi^{(0)}(r)
\Bigr].
\label{33a}
\end{eqnarray}
In order to prove this relation, we represent Eq.~(\ref{u0bare}) in the form
\begin{eqnarray}
{U}^{(0)}(0)
=\int\d^3r\Bigl[\frac{\hbar^2}{m}|\nabla\psi^{(0)}(r)|^2
                            +[\varphi^{(0)}(r)]^2 \Phi(r)\Bigr],
\label{33b}
\end{eqnarray}
which can be found using integration by parts and taking into account the
Schr\"odinger equation (\ref{twobody}) and the boundary condition
(\ref{scatasymp}).  Further, varying  Eq.~(\ref{33b}) and keeping in mind
Eqs.~(\ref{twobody}) and (\ref{scatasymp}), we arrive at Eq.~(\ref{33a}). The relation~(\ref{33a})
bears analogy to the variational theorem for the energy:
\[
\delta E_{n}=\int\d^{3}r\,\varphi^{(0)}_{n}({\bf r})
\delta\Bigl(-\frac{\hbar^2}{m}\nabla^2 +\Phi(r)\Bigr)
\varphi^{(0)}_{n}({\bf r}),
\]
here the real wave function $\varphi^{(0)}_{n}({\bf r})$ obeys the
Schr\"o\-din\-ger equation for a bound state with the energy $E_{n}<0$.
Equation (\ref{33b}) can be represented in the more convenient form
\begin{equation}
\int\d^3r\,[\varphi^{(0)}(r)]^2\Phi(r)=4\pi\hbar^2(a-b)/m.
\label{31int}
\end{equation}
Here one more characteristic length $b$ (in addition to $a$) has been
introduced:
\begin{eqnarray}
b=\frac{1}{4\pi}\int\d^{3}r\,\bigl|\nabla\psi^{(0)}(r)\bigr|^{2}.
\label{22b}
\end{eqnarray}
It follows from Eq.~(\ref{22b}) that $b$ is a positive quantity. We stress
that $b$ is not expressed in terms of $a$ and depends on a particular shape
of the interaction potential $\Phi(r)$.  For example, when $\Phi(r)$ is the
hard-sphere potential~(\ref{hardsph}), we have $b=a$. While for $\Phi(r)$
close to zero, in the weak-coupling regime, we have $b\simeq -a_{1}$,
$a\simeq a_{0}$, and, hence, $b\ll a$.

Lastly, from the definitions (\ref{adef}) and (\ref{22b}) and the variational
theorem (\ref{33a}) it follows that
\begin{equation}
\gamma\frac{\partial a}{\partial \gamma}=
m\frac{\partial a}{\partial m}=a-b,
\label{33}
\end{equation}
where we introduce the auxiliary parameter $\gamma$ called coupling constant
[i.e. $\Phi(r)\to\gamma\Phi(r)$].  The first equality in Eq.~(\ref{33})
implies that $a=a(\gamma m)$, which is an obvious consequence of the
definition (\ref{adef}) and the Schr\"odinger equation (\ref{twobody}). The
relations (\ref{33}) demonstrate that the quantity $b$ is expressed in terms
of $a$ and its derivative with respect to $\gamma$ (or $m$) rather than in
terms of $a$.

\section{Variational treatment of the Bogoliubov model}
\label{2sec}

Although the aim of this paper is to investigate a dilute Bose gas with
the strong-coupling interaction, it is deductive to start with the
Bogoliubov model related to the weak-coupling regime. This regime
implies a minor role of particle scattering, both in medium and
out of it, and, thus, is characterized by the following inequalities
for the scattering waves (\ref{4}):
\begin{equation}
|\psi(r)| \ll 1, \;\quad |\psi_{{\bf p}}({\bf r})| \ll 1.
\label{9}
\end{equation}
In particular, the Bogoliubov model operates with the
choice~\cite{Ch1,Ch2,Ch3}
\begin{eqnarray}
|\psi(r)| \ll 1, \;\quad \psi_{{\bf p}}({\bf r}) = 0.
\nonumber
\end{eqnarray}
As the depletion of the Bose condensate $(n-n_0)/n$ is small in
a weakly interacting many-boson system,  we have for the one-particle
density matrix $F_{1}(r)=\langle\hat{\psi}^{\dagger}({\bf r}_{1})
\hat{\psi}({\bf r}_{2})\rangle$:
$$
\left|\frac{F_{1}(r)}{n}\right|=
\left|\int\frac{\d^3k}{(2\pi)^3}\frac{n_k}{n}
\exp(i{\bf k}\cdot{\bf r})\right|\leq \frac{n-n_{0}}{n}\ll 1.
$$
So, the Bogoliubov scheme of treating a Bose gas involves two
small quantities $\psi(r)$ and $F_{1}(r)/n$, and completely neglects
scattering in the supracondensate-condensate sector of $g(r)$:
$\psi_{{\bf p}}({\bf r})=0$.  This along with Eq.~(\ref{3}) allows for
rewriting Eq.~(\ref{8}) in the following form:
\begin{equation}
g(r)=1+2\psi(r)+\frac{2}{n}
\int\frac{\d^3k}{(2\pi)^3}n_k\exp(i{\bf k}\cdot{\bf r}).
\label{11}
\end{equation}
Here we have restricted ourselves to the terms linear in $\psi(r)$
and $F_{1}(r)/n$. Besides, it is implied that $\psi^*(r)=\psi(r)$,
for the pair wave functions can be chosen as real quantities.
Inserting Eq.~(\ref{11}) into Eq.~(\ref{7}), we are able to employ
a variational procedure to derive the unknown quantities ${\psi}(k)$
and $n_k$. In so doing, we should realize that $n_k$ and ${\psi}(k)$
are not independent variables. In fact, there are no spatial boson
correlations in absence of interaction~\cite{Not3a}. Hence, in this
case ${\psi}(k)=0$, and, as we investigate the ground state, all the
bosons are condensed, $n_k=0$. While in presence of interaction
${\psi}(k)\not=0$, which leads to a nonzero depletion and, so,
$n_k\not=0$. Within the Bogoliubov model ${\psi}(k)$ is related to
$n_k$ by
\begin{equation}
n_k (n_k+1)=n_0^2{\psi}^2(k).
\label{bogrel}
\end{equation}
Indeed, according to the canonical Bogoliubov transformation,
quasiparticle (bogolon) operators $\hat{\alpha}^{\dagger}_{{\bf k}}$ and
$\hat{\alpha}_{{\bf k}}$ are connected with the operators of the
primordial bosons by the expression
\begin{equation}
\hat{a}_{{\bf k}}=u_{k}\hat{\alpha}_{{\bf k}}+v_{k}
\hat{\alpha}^{\dagger}_{-{\bf k}},\quad
\hat{a}^{\dagger}_{{\bf k}}=u_{k}\hat{\alpha}^{\dagger}_{{\bf k}}+
v_{k}\hat{\alpha}_{-{\bf k}},
\label{bogtrans}
\end{equation}
where
\begin{equation}
u_{k}^2-v_{k}^2=1.
\label{13}
\end{equation}
Within the Bogoliubov model the ground state of the system of
interest is the bogolon vacuum, and, so, at the zero
temperature we have
\begin{equation}
\langle\hat{\alpha}^{\dagger}_{{\bf k}} \hat{ \alpha}_{{\bf k}}
\rangle=0.
\label{ground}\end{equation}
Then, using (\ref{25}), (\ref{bogtrans}) and (\ref{ground}), one
can find
\[
n_k=v_{k}^2, \quad
      {\psi}(k)=u_{k}v_{k}/n_0,
\]
which in conjunction with (\ref{13}) leads to Eq. (\ref{bogrel}).  We remark
that beyond the Bogoliubov scheme Eq.~(\ref{bogrel}) is not valid and should
be corrected (see Eq.~(\ref{rel}) in the present paper and discussion on this
question in Ref.~\cite{Ch1}).

Now, inserting Eq. (\ref{11}) into Eq. (\ref{7}) and varying the
obtained expression with respect to $\psi(k)$ and $n_k$, we derive
\begin{equation}
\delta\varepsilon=\int\frac{\d^3k}{(2\pi)^3}
\Bigl[
\Bigl(T_k+n\Phi(k)\Bigr)\frac{\delta n_k}{n} + n\Phi(k)\delta\psi(k)
\Bigr].
\label{bogvar}
\end{equation}
According to Eq.~(\ref{bogrel}) infinitesimal changes $\delta\psi(k)$
and $\delta n_k$ are connected by
\begin{equation}
\delta{\psi}(k)=\frac{(2n_k+1)\delta n_k}{2n_0^2{\psi}(k)}
+\frac{{\psi}(k)}{n_0}\int\frac{\d^3q}{(2\pi)^3}\delta n_q,
\label{varpsi}
\end{equation}
where the equality
$$
n=n_0+\int\frac{\d^3k}{(2\pi)^3}n_k
$$
is implied. Taking $\delta\varepsilon=0$ and using Eqs.~(\ref{bogvar})
and (\ref{varpsi}), we find the following equation:
\begin{eqnarray}
-2T_k{\psi}(k)=\frac{n^2}{n^2_0}
&&{\Phi}(k)(1+2n_k)+2n{\psi}(k)
\nonumber\\
&&\times\left({\Phi}(k)+\frac{n}{n_0}
\int\frac{\d^3q}{(2\pi)^3}{\Phi}(q){\psi}(q)\right).
\label{18}
\end{eqnarray}
Note that Eq.~(\ref{18}) is able to yield results being accurate
only to the leading order in $(n-n_0)/n$ because Eq. (\ref{11}) is
valid to the next-to-leading order~\cite{Not0}. So, Eq.~(\ref{18})
should be written in the form
\begin{equation}
-2T_k {\psi}(k)={\Phi}(k)(1+2n_k)+2n{\psi}(k)\Phi(k),
\label{bgbog}
\end{equation}
which, with the help of Eq.~(\ref{11}), can be represented as
\begin{equation}
\frac{\hbar^2}{m}\nabla^2\varphi(r)=\Phi(r)
+n\int \d^3y\,\Phi(y)\bigl(g(|{\bf r}-{\bf y}|)-1\bigr).
\label{20}
\end{equation}
Equation~(\ref{20}) is very similar to the Bethe-Goldstone one.
Necessary details concerning Eq.~(\ref{20}) can be found in the
papers~\cite{Ch3,Shan1}. We remark that the right-hand side (r.h.s.)
of Eq.~(\ref{20}) can be thought of as the in-medium potential of the
boson-boson interaction in the Born approximation. Indeed,
Eq.~(\ref{20}) is derived from the more general equation given by
Eq.~(\ref{11e}) by means of linearization in $\psi(r)$ and
$F_{1}(r)/n$. So, the Bogoliubov model can be treated as the
in-medium Born approximation for its generalization developed on the
basis of Eqs.~(\ref{3}) and (\ref{8}) beyond the weak-coupling regime
(see section \ref{genBog}). In accordance with this treatment,
Eq.~(\ref{20}) at $n=0$ is nothing but the Fourier transform of
Eq.~(\ref{bornpsi}) while Eq.~(\ref{11e}) is reduced to the exact
Schr\"odinger equation (\ref{twobody}) at $n=0$. We will return to
this important point in Sec.~\ref{effpot}.

The system of Eqs.~(\ref{bogrel}) (here we should set $n=n_0$) and
(\ref{bgbog}) can easily be solved, which leads to the familiar
results~\cite{Bog1}:
\begin{eqnarray}
n_k
&=&\frac{1}{2}\left(\frac{T_k+n{\Phi}(k)}{\sqrt{T_k^2+
2nT_k{\Phi}(k)}}-1\right),
\label{21aa}\\
{\psi}(k)
&=&-\frac{1}{2}\frac{{\Phi}(k)}{\sqrt{T_k^2+2nT_k{\Phi}(k)}}\,.
\label{psibog}
\end{eqnarray}

\section{Density expansions in the Bogoliubov model}
\label{densbog}

As it was mentioned, in this paper we investigate the strong-coupling
regime for a dilute Bose gas. So, considering a dilute Bose gas in
the weak-coupling approximation can be a good exercise providing us
with useful information. Let us investigate the thermodynamics of a
dilute many-boson system within the Bogoliubov model. With
Eqs.~(\ref{7}), (\ref{11}), (\ref{21aa}) and (\ref{psibog}) we derive
\begin{eqnarray}
\varepsilon
=\frac{n}{2}{\Phi}(0)
+&&\frac{1}{2n}\int\frac{\d^3q}{(2\pi)^3}
\nonumber\\
&&\times\Bigl(\sqrt{T_q^2+2nT_q{\Phi}(q)}-T_q - n{\Phi}(q)\Bigr).
\label{21a}
\end{eqnarray}
Here we describe in details the method for obtaining the low-density
expansions for expressions like Eq.~(\ref{21a}). This equation can be
represented in the following form:
\begin{eqnarray}
\varepsilon=\frac{n}{2}\left({\Phi}(0)
-\int\frac{\d^3q}{(2\pi)^3}\frac{{\Phi}^2(q)}{2T_q}\right)+I,
\label{21b}
\end{eqnarray}
where
\begin{eqnarray}
I=\frac{1}{2}\int&&\frac{\d^{3}q}{(2\pi)^3}
\nonumber \\
&&\times\biggl(\sqrt{\frac{T_q^2}{n^{2}}+2\frac{T_q}{n}{\Phi}(q)}
-\frac{T_q}{n} - {\Phi}(q)+\frac{{\Phi}^2(q)}{2(T_q/n)}\biggr).
\nonumber
\end{eqnarray}
Now, with the ``scaling" substitution
\begin{equation}
{\bf q}={\bf q'}\sqrt{2mn}/\hbar
\label{subst}
\end{equation}
we derive
\begin{equation}
\frac{I}{n^{3/2}}=\frac{1}{2}\left(\frac{2m}{\hbar^{2}}\right)^{3/2}
\int\frac{\d^{3}q'}{(2\pi)^{3}}f(q',n),
\label{In}
\end{equation}
where
\begin{eqnarray}
f(q',n)
&=&\sqrt{(q')^{4}+2(q')^{2}\Phi(q'\sqrt{2mn}/\hbar)}-(q')^{2}
\nonumber \\
&&-\Phi(q'\sqrt{2mn}/\hbar)+\frac{\Phi^{2}(q'\sqrt{2mn}/\hbar)}{2(q')^{2}}.
\nonumber
\end{eqnarray}
The advantage of the representation (\ref{21b}) is that the resulted integral
in Eq.~(\ref{In}) uniformly converges for $q'\to\infty$ with respect to $n$
for $n\to 0$, and, thus, we obtain
\begin{eqnarray}
\lim_{n\to 0}\frac{I}{n^{3/2}}
=\frac{1}{2}\left(\frac{2m}{\hbar^{2}}\right)^{3/2}
\int\frac{\d^{3}q'}{(2\pi)^{3}}f(q',n=0).
\nonumber
\end{eqnarray}
Here the integral is readily calculated, and the main asymptotics for $I(n)$
is given by
\begin{equation}
I\simeq\frac{8}{15\pi^2} \left(\frac{m n}{\hbar^2}\right)^{3/2}
{\Phi}^{5/2}(0).
\label{21e}
\end{equation}
Further, with the help of Eqs.~(\ref{borna}), (\ref{21b}) and (\ref{21e}),
the expression (\ref{21a}) is rewritten as
\begin{equation}
\varepsilon=\frac{2\pi\hbar^2 n(a_{0}+a_{1})}{m}
+\frac{2\pi\hbar^2 n a_{0}}{m}\frac{128}{15\sqrt{\pi}}\sqrt{n a_{0}^3}
+\cdots.
\label{21i}
\end{equation}
Thus, we obtain the first two terms in the density expansion for the energy
per particle within the Bogoliubov model.

The density expansion for the condensate depletion is inferred
from Eq.~(\ref{21aa}) in the same manner employing the substitution
(\ref{subst}):
\begin{equation}
\zeta=\frac{n-n_0}{n}=\int \frac{\d^{3}q}{(2\pi)^{3}}\frac{n_{q}}{n}
=\frac{8\sqrt{na_{0}^3}}{3\sqrt{\pi}}+\cdots.
\label{bogdepl}
\end{equation}

Now let us discuss the range of validity of the Bogoliubov model. First, the
condensate depletion (\ref{bogdepl}) should be small as the representation
for the pair correlation function (\ref{3}) which we start from is valid only
in this case. Note that if we expand the depletion with respect to the
coupling constant $\gamma$ (we assume that $\Phi(r) \propto \gamma$), we also
arrive at Eq.~(\ref{bogdepl}) since the occupation number (\ref{21aa})
depends only on the production $n\gamma$. Thus, the condition $na_{0}^{3}\ll
1$ should be fulfilled. Second, we exploit the weak-coupling character of the
pairwise interaction $\Phi(r)$, which implies that the condition
(\ref{borncrit}) should also be satisfied. Note that Bogoliubov himself
realized this necessary condition since he treated the term
$n{\Phi}(0)/2=2\pi\hbar^2 na_{0}/m$ involved in the mean energy (\ref{21a})
as the major one~\cite{Bog1}. Beyond the inequality (\ref{borncrit}) the
model may be thermodynamically unstable. In particular, the opposite case
\begin{eqnarray}
{\Phi}(0) < \int\frac{\d^3k}{(2\pi)^3}\frac{{\Phi}^2(k)}{2T_k}
\label{opp}
\end{eqnarray}
leads to the negative scattering length in the next-to-Born
approximation $a=a_{0}+a_{1}<0$ which at sufficiently low densities
results in the incorrect sign for the compressibility
$-\partial^{2}E/\partial V^{2}=\partial P/\partial V > 0$ as it is
seen from Eq.~(\ref{21i}). Note that this important point is not
always stressed in literature. Moreover, Bru and Zagrebnov~\cite{bru}
proposed the model reduced to the Bogoliubov approach but in
conjunction with the inequality (\ref{opp}). We have to conclude that
this model hardly has a physical sense.

\section{Short-range correlations and ultraviolet divergence
within the effective-interaction approach}
\label{effpot}

After the detailed consideration fulfilled in the previous sections,
we can argue that the Bogoliubov model with the ``bare" potential
$\Phi(r)$ replaced by an effective one ~\cite{Lee}, is
thermodynamically inconsistent. Indeed, the basic relations of the
Bogoliubov model~(\ref{11}) and (\ref{bogrel}) depend on interaction
implicitly. Hence, the pairwise potential appearing in
Eqs.~(\ref{bgbog})-(\ref{psibog}) comes from Eq.~(\ref{7}) that is the
general relation valid in both the weak- and strong-coupling regimes.
Thus, a calculating procedure based on (\ref{11}) and
(\ref{bogrel}) has to eventuate in Eqs.~(\ref{bgbog})-(\ref{psibog}).
Otherwise, like in the case of using the Bogoliubov model with an
effective interaction, this procedure does not yield the result
minimizing the mean energy. Note that we do not mean, of course, that
the t-matrix approach or the pseudopotential method can not be
applied in the quantum scattering problem. It is only stated that the
usual way of combining the ladder diagrams with the RPA ones
(``bubbles") leads to the thermodynamic inconsistency.

To clarify the reason for this inconsistency, let us take a look at
the picture of the spatial boson correlations derived in the
framework of the effective-interaction approach. According to the
paper by Hugenholtz and Pines~\cite{Lee} (see Eq. (5.10a) therein),
the structural factor
\begin{equation}
S(k)=1+n\int\d^3r\,[g(r)-1]\exp(-i{\bf k}\cdot{\bf r})
\label{strdef}
\end{equation}
of a strong-coupling Bose gas can be written at $n \to 0$ as follows:
\begin{equation}
S(k)=1+2\frac{n_0}{n}\langle \hat{a}_{{\bf k}}^{\dagger}
\hat{a}_{{\bf k}}\rangle
+\frac{n_0}{n}\bigl(\langle \hat{a}_{{\bf k}}\hat{a}_{-{\bf k}}
\rangle
+\langle \hat{a}_{-{\bf k}}^{\dagger}\hat{a}_{{\bf k}}^{\dagger}
\rangle\bigr).
\label{1}
\end{equation}
Using Eq.~(\ref{25}), the equality ${\psi}(k)=
{\psi}^*(k)$~\cite{Not1a} and definition (\ref{strdef}) of the
structural factor, one can readily verify that Eq.~(\ref{1}) is
reduced to Eq.~(\ref{11}). This Bogoliubov relation does not depend
on the interaction potential explicitly. So, use of the ``dressed"
interaction can in no way disturb the form of Eq.~(\ref{11}), and,
therefore, the effective-interaction approach deals with the pair
distribution function whose structure has the obvious weak-coupling
character. In particular, from Eq.~(\ref{1}) it can be found that $g(r)
\to 1+2\psi^{(0)}(r)$ for $n \to 0$ as opposed to the correct
strong-coupling result given by Eq.~(\ref{5}). However, the wave
function $\varphi^{(0)}(r)$ obeys Eq.~(\ref{twobody}) in the
effective-interaction approach~\cite{Not2}. While within the
Bogoliubov model $\varphi^{(0)}(r)$ is the solution of Eq.~(\ref{20})
at $n=0$. This equation (Eq.~(\ref{20}) at $n=0$) comes from the
Schr\"odinger equation~(\ref{twobody}) in the Born approximation~(see
discussion in Sec.~\ref{2sec}). Thus, the effective-interaction
approach is not totally reduced to the weak-coupling framework due to
its features of the strong-coupling character. {\it Exactly the
combination of the peculiarities of both the strong- and
weak-coupling regimes is the reason for the thermodynamic
inconsistency mentioned above.}

It is also worth noting that this combination of the features of
weakly and strongly interacting Bose gases leads not only to the
thermodynamic inconsistency. It results also in one more
misrepresentation of the effective-interaction approach. We mean an
irrelevant picture of the short-range boson correlations. Indeed, in
the case of a strongly singular pair interaction for the solution of
Eq.~(\ref{twobody}) we have $\varphi^{(0)}(r=0)=0$~[see
Sec.~\ref{1asec}], which provides $\psi^{(0)}(r=0)=-1$. Within the
effective-interaction scheme $g(r)$ obeys Eq.~(\ref{11}) while
$\varphi^{(0)}(r)$ satisfies Eq.~(\ref{twobody}). This implies that
$g(r=0)\to 1+2\psi^{(0)}(r=0)=-1$ in the zero density limit when
$(n-n_0)/n \to 0.$ The obtained result does not agree with the
physical sense of $g(r)$ (the conditional probability) and has
nothing to do with the strong-coupling regime when the relation
$g(r=0)=0$ has to be satisfied. The situation even aggravates if we
recall that the scattering parts of the supracondensate-condensate
pair wave functions $\psi_{\bf p}({\bf r})$ are equal to zero in the
Bogoliubov model. So, in what concerns the pair distribution
function, the ``triple" correlations involved in Eq.~(\ref{25}) for
$\psi_{\bf p}({\bf k})$ are completely ignored within the
effective-interaction scheme. However, when deriving an equation for
the effective potential, these correlations are taken into
consideration, for example, within the Beliaev approach (see
discussion in the review ~\cite{shigrif}). So, we face one more
combination of the weak- and strong-coupling features being
characteristic of the approach of Ref.~\cite{Lee}.

Since the short-range behaviour of the pair distribution function is
not correct within the effective-interaction approach, one can expect
some problems related to evaluation of the mean energy (\ref{7}). Let
us consider the effective-interaction method in its simplest variant,
the so-called pseudopotential model (see paper by Lee, Huang and Yang
of Ref.~\cite{Lee}). This variant implies the replacement $\Phi(r)\to
\delta({\bf r})4\pi\hbar^{2}a/m$, and, hence, for the Fourier
transform we have
\begin{equation}
\Phi(k) \to 4\pi\hbar^{2}a/m=\mbox{const}, \label{hardsp}
\end{equation}
where $a$ is the scattering length (\ref{adef}) obtained from the
Schr\"odinger equation (\ref{twobody}). So, the pseudopotential model
is reduced to the Bogoliubov one with the effective pairwise
interaction given by (\ref{hardsp}). In the well-known
textbook~\cite{Fet} one can find two ways of calculating the leading
and next-to-leading terms of the low-density expansion of the energy
of a dilute Bose gas within the pseudopotential model. One of them,
see the pages 314-319, consists in dealing directly with the
Hamiltonian of the system and faces the divergent integral $\int
\d^3k/k^2$~(the ultraviolet divergence). The latter, given on the
pages 218-223, allows for calculating the difference
$\varepsilon-\mu/2$ and does not lead to any divergence.

In the previous section we have derived the low-density expansion
(\ref{21i}) corresponding to the Bogoliubov model. This expansion can
help us to understand reasons for the ambiguous result of the
pseudopotential model. Use of the pseudopotential (\ref{hardsp})
leads to the substitution $a\to a_{0}$ in Eq.~(\ref{21i}). Besides,
$a_{1}\to -\infty$ as it becomes proportional to $\int \d^3k/k^2$
[see Eqs.~(\ref{borna})]. This agrees with the result of the first
way of calculating $\varepsilon$ in the textbook~\cite{Fet}. The
divergent integral is usually removed  because it is assumed that
``this divergence is not very basic"~\cite{Fet}. So, we arrive at the
correct expression (\ref{23}) which is found in our model beyond any
divergences. The reason for the singularity is obvious because the
necessary condition (\ref{borncrit}) of the validity of the
Bogoliubov model is not satisfied. However, the question remains why
the pseudopotential approach results nevertheless in the correct
final expression (\ref{23})? The point is that the
effective-interaction scheme actually involves an additional
assumption, namely, the Landau postulate (see the footnote in
Ref.~\cite{Bog1} and discussion in Ref.~\cite{Lee}). This postulate
asserts that the properties of dilute quantum gases are ruled by the
scattering length $a$~\cite{Not4}. Let us consider how the additional
assumption is used when deriving the low-density expansion for the
mean energy. According to the Landau argument this expansion should
be of the form
\begin{equation}
\varepsilon=c_{1}(a)n + c_{2}(a)n^{3/2}+\cdots,
\label{enexp}
\end{equation}
where the factors $c_{i}$ can depend on various quantities but one of
them, the ``bare" potential $\Phi(r)$, is involved only through the
mediation of the scattering length (\ref{adef}). Substituting the
Born series (\ref{expa}) in the expressions for $c_{i}(a)$ in the
{\it weak-coupling} regime, we obtain
\begin{eqnarray}
c_{i}(a)&=&c_{i}(a_{0}+a_{1}+\cdots)
\nonumber \\
&\simeq& c_{i}(a_{0})+
\left.\frac{\partial c_{i}(a)}{\partial a}
\right|_{a=a_{0}}a_{1}+\cdots.
\label{enexp1}
\end{eqnarray}
As the functional dependencies $c_{i}(a)$ are of the same form in
both the weak- and strong-coupling regimes, one is able to restore
them by keeping the Born terms $c_{i}(a_{0})$ in the expansion
(\ref{enexp}) and omitting others (dependent on
$a_{1},\,a_{2},\,\cdots$). It can readily be verified in this way that
Eq.~(\ref{21i}) leads to Eq.~(\ref{23}). Thus, the pseudopotential approach provides
(after the regularization) the correct result given by Eq.~(\ref{23})
because it is equivalent, in the first two orders of the low-density
expansion, to the calculating scheme using the Bogoliubov model
together with Eqs.~(\ref{enexp}) and (\ref{enexp1}) based on the
Landau postulate~\cite{Note5a}. Note that this simple scheme looks even more accurate
and justified than the pseudopotential approach. At least, it allows for
investigating a strongly interacting Bose gas beyond any ultraviolet
divergence which appears as a result of violating subtle balance of
the correlation terms coming from the boson-boson scattering.
However, both the pseudopotential approach and the Bogoliubov model used
together with Eqs.~(\ref{enexp}) and (\ref{enexp1}) cannot yield adequate
microscopical results concerning the strong-coupling regime.

The second way of calculating $\varepsilon(n)$ within the
pseudopotential model allows one to find the low-density expansion
(\ref{enexp}) starting from the difference
\begin{equation}
\varepsilon-\frac{\mu}{2}=\varepsilon-\frac{1}{2}
\frac{\partial (\varepsilon n)}{\partial n}\simeq
-\frac{2\pi\hbar^2 n a_{0}}{m}\frac{32}{15\sqrt{\pi}}\sqrt{n a_{0}^3},
\label{2way}
\end{equation}
where Eq.~(\ref{21i}) and the well-known thermodynamic relation
$\mu=\partial \bigl(n\varepsilon(n)\bigr)/\partial n$ are of use. There is no
divergent integral here due to the specific property of the expansion
(\ref{21i}): $a_{1}$ is only involved in the leading-order term being
exactly cancelled in Eq.~(\ref{2way})~\cite{Note5a}. The solution of the
differential equation (\ref{2way}) (after replacing $a_0$ by $a$) is
of the form $\varepsilon=c_{1}n+\dfrac{2\pi\hbar^2 n a}{m}\dfrac{128}
{15\sqrt{\pi}}\sqrt{n a^3}$. To specify $c_{1}$ being a constant of
integration, one again needs to involve information additional to
Eq.~(\ref{2way}). Following Landau and {\it postulating} that $c_{1}$
depends on the pairwise potential only through the mediation of the
scattering length $a$, one arrives at $c_1 =2\pi\hbar^2 a/m$, which
eventuates in Eq.~(\ref{23}).

Thus, we remark one more that the effective-interaction approach {\it
taken in conjunction with the Landau postulate} yields the correct
expansion (\ref{23}). Even the Wu's term~\cite{Wu} in the low-density
expansion of the energy of a strong-coupling Bose gas is likely to be
correct because it is present in the weak-coupling calculations
beyond the Bogoliubov model~\cite{Tolmach}. However, the
microscopical results found within the effective-interaction approach
should be reexamined. So, a correct strong-coupling generalization of
the Bogoliubov model should be constructed. It is also of importance
that the density expansions for the quantities depending on the form of
$\Phi(r)$, for example, the interaction (\ref{intdef}) and kinetic
(\ref{kindef}) energies, cannot directly be derived within the
effective-interaction scheme. We discuss this point in Sec.~\ref{intkinsec}.

\section{Strong-coupling generalization of the Bogoliubov model}
\label{genBog}

To avoid the serious problems mentioned in the previous section, we
should abondon the effective-interaction method and use a way based on
Eq.~(\ref{3}). Equations (\ref{3}), (\ref{8}) and (\ref{7}) make it
possible to express $\varepsilon$ in terms of the pair wave functions
and momentum distribution. So, the variational procedure similar to
that of Sec.~\ref{2sec} can be employed to determine these basic
quantities. In so doing we should again keep in mind that the
momentum distribution depends on the scattering waves~(see
Sec.~\ref{2sec}). However, now we are not able to use the Bogoliubov
relation (\ref{bogrel}) which does not take into account scattering
in the supracondensate-condensate sector. In the paper~\cite{Ch1} the
following extension of Eq.~(\ref{bogrel}) has been proposed:
\begin{equation}
n_k(n_k+1)= n_0^2{\psi}^2(k)+ 2n_0\int\frac{\d^3q}{(2\pi)^3}n_q
                                      {\psi}^2_{{\bf q}/2}({\bf k}).
\label{rel}
\end{equation}
This expression has been derived with the help of the reasonable
expectation that the equation for $\psi_{{\bf p}}({\bf k})$ should be
reduced to the equation for $\psi(k)$ in the limit $p \to 0.$ It is
interesting to note the obvious structural parallels between
Eqs.~(\ref{3}) and (\ref{rel}). Now, inserting Eqs.~(\ref{3}) and
(\ref{8}) into Eq.~(\ref{7}) and, then, perturbing ${\psi}(k)$ and
$n_k$ under the condition (\ref{rel}), we find:
\begin{equation}
-2 T_k {\psi}(k)={U}(k)(1+2 n_k)+2n{\psi}(k){U}'(k).
\label{11a}
\end{equation}
Here ${U}(k)$ and $U'(k)$ are defined by
\begin{eqnarray}
{U}(k)&=&\int\d^3r\,\varphi(r)\Phi(r)\exp(-i{\bf k}\cdot{\bf r}),
\label{11b}\\
 U'(k) &=&\int\d^3r\,\bigl(\varphi_{{\bf k}/2}^2({\bf r})-
     \varphi^2(r)\bigr)\Phi(r)
\nonumber \\ &&-\int\frac{\d^3q}{(2\pi)^3}\frac{{U}(q)}{{\psi}(q)}
     \bigl({\psi}^2_{{\bf k}/2}({\bf q})-
                {\psi}^2(q)\bigr).
\label{11c}
\end{eqnarray}
An equation for $\psi_{{\bf p}}({\bf k})$ can be derived in the same
manner. Note that we have the following limiting relation:
\begin{equation}
\lim_{p \to 0}\varphi_{{\bf p}}({\bf r})=\sqrt{2}\varphi(r),
\label{limphip}
\end{equation}
where the factor $\sqrt{2}$ comes from the second expression in
Eq.~(\ref{4}). Using the equation for $\psi_{{\bf p}}({\bf k})$, one
can be convinced that $\psi_0(k)\equiv\lim_{p\to 0}\psi_{{\bf
p}}({\bf k})= \mbox{const}\cdot\psi(k)$. We should put this constant
equal to $\sqrt{2}$ in order to obtain
Eq.~(\ref{limphip})~\cite{Note3}. So, we have
$\varphi(r)-\sqrt{2}\varphi_{{\bf p}}({\bf r}) \propto p^2$~(see Ref.
10 in the paper \cite{Ch1}), which provides $U'(k)-U(k) \propto k^4$
for $k \to 0$. Besides, it is easy to verify that $[U(k)-U'(k)]/T_k
\to 0$ when $k \to \infty$. Therefore at small densities
$n[U'(k)-U(k)]\ll T_k$ for all momenta. This is why the difference
between $U'(k)$ and $U(k)$ does not play a role when calculating the
first two terms of the low-density expansions for the basic
thermodynamic quantities. Thus, at sufficiently small densities
Eq.~(\ref{11a}) can be rewritten in the following form:
\begin{eqnarray}
\frac{\hbar^2}{m}\nabla^2\varphi(r)
&=&\varphi(r)\Phi(r)
\nonumber \\
&&+n\int\d^3y\,\varphi(y)\Phi(y)\bigl(g_{tr}(|{\bf r}-{\bf y}|)-1\bigr),
\label{11e}
\end{eqnarray}
where $g_{tr}(r)$ stands for the truncated pair distribution function
which is equal to the right-hand side of Eq.~(\ref{11}) even beyond
the weak-coupling regime.

Equations (\ref{3}) [and, hence, Eqs.~(\ref{8}), (\ref{7})] and (\ref{rel}) are written with the
con\-den\-sate-con\-den\-sate and
sup\-ra\-con\-den\-sate-con\-den\-sate ``channels" taken into
account. As these relations are accurate to the next-to-leading order
in $(n-n_0)/n$, then, Eq.~(\ref{11a}) can be accurate only to the
leading order in $(n-n_0)/n$. So, it would be wrong to solve
Eq.~(\ref{11a}) together with Eq.~(\ref{rel}). One should investigate
Eq.~(\ref{11a}) in conjunction with the shortened version of
Eq.~(\ref{rel}) given by Eq.~(\ref{bogrel}) where the equality
$n=n_0$ is implied. The system of Eqs.~(\ref{bogrel}) and (\ref{11a})
has the following solution:
\begin{eqnarray}
    n_k&=&
\frac{1}{2}\Biggl(\frac{\widetilde{T}_k+nU(k)}
{\sqrt{\widetilde{T}_k^2+2n\widetilde{T}_k U(k)}} -1\Biggr),
\label{12a}
\\ \psi(k)&=&
-\frac{1}{2}\frac{U(k)}{\sqrt{\widetilde{T}_k^2+2n\widetilde{T}_k
U(k)}} \label{12}
\end{eqnarray}
with $\widetilde{T}_k =T_k+n[U'(k)-U(k)]$. As to the
sup\-ra\-con\-den\-sate-condensate states, the goal of this paper
makes it possible not to go into details concerning $\varphi_{{\bf
p}}({\bf r})$. It is sufficient only to use Eq.~(\ref{limphip}). In
the zero-density limit $n=0$ Eq.~(\ref{12}) is reduced to the
equation ${\psi}(k)=\psi^{(0)}(k)= -{U}^{(0)}(k) /(2T_k)$, which can
be rewritten in the form of Eq.~(\ref{twobody}). So, at small
densities all thermodynamic quantities can be expressed in terms of
the vacuum (out-medium) scattering amplitude ${U}^{(0)}(k)$ given by
\begin{equation}
{U}^{(0)}(k)=\int\d^3r\,
\varphi^{(0)}(r)\Phi(r)\exp(-i{\bf k}\cdot{\bf r}).
\label{U0}
\end{equation}
Below it is shown that this feature of Eqs.~(\ref{12a}) and
(\ref{12}) results directly in Eqs.~(\ref{depletion}), (\ref{26b})
and (\ref{23}). So, the low-density energy expansion given by
Eq.~(\ref{23}) is determined in our model beyond any additional
assumptions like the Landau postulate in the pseudopotential
approach. Equations~(\ref{12a}) and (\ref{12}) yield $n_k \propto
1/k$ and $\psi(k) \propto 1/k$ at small boson momenta.  This is
totally consistent with the well-known ``$1/k^2$" Bogoliubov
theorem~\cite{Bog2}. It is interesting that the correct low-momentum
behaviour of $n_k$ and $\psi(k)$ comes from the relation
$U'(k)-U(k)\propto k^4$ which follows from Eq.~(\ref{11c}) taken in
conjunction with Eq.~(\ref{limphip}) being a result of the principle
of correlation weakening. Note that Eqs.~(\ref{21aa}) and
(\ref{psibog}) derived within the Bogoliubov model can be obtained
from Eqs.~(\ref{12a}) and (\ref{12}) by replacing $\widetilde{T}_k$
and ${U}(k)$ by $T_k$ and ${\Phi}(k)$, respectively. So, in what
concerns Eqs.~(\ref{12a}) and (\ref{12}), the situation in our
strong-coupling generalization of the Bogoliubov model does look as
if we operated with a weakly interacting Bose gas of the
quasiparticles with the renormalized kinetic energy $\widetilde{T}_k$
and the effective interaction $U(r)=\varphi(r)\Phi(r)$. This is close
to the expectations following from the effective-interaction
approach~\cite{Lee}. Due to the boundary condition $U(k)-U'(k)\propto
k^4$ at small $k$ the mass of the quasiparticles coincides with that
of the primordial bosons. However, as it has been mentioned in
Ref.~\cite{Ch1}, Eq.~(\ref{rel}) is the simplest of the possible
approximations which are fixed by the necessary condition that the
Bethe-Goldstone equation for ${\psi}_{{\bf p}}({\bf k})$ is reduced
to the equation for ${\psi}(k)$ in the limit $p \to 0.$ These
approximations lead to the same low-density expansions for
thermodynamic quantities (the energy, the chemical potential, the
condensate depletion) in the leading and next-to-leading terms.
However, difference in the thermodynamics has to appear at small but
finite densities as well as in the higher-order terms in the density
expansions. The same goes for the microscopical picture given by
Eqs.~(\ref{12a}) and (\ref{12}). Here the different approximations of
the relationship between the momentum distribution and pair wave
functions can lead to some different details. In particular, we can
not {\it a priori} exclude that there is a more relevant variant of
Eq.~(\ref{rel}) which leads to $U'(k)-U(k) \propto k^2\;(k \to 0)$.
In this situation the quasiparticle mass would be $m^*=m/(1+\beta
n)$, where $\beta =\lim_{k \to 0} [U'(k)-U(k)]/T_k$.

\section{Short-range boson spatial correlations}
\label{4sec}

Now, to elaborate on the picture of the short-range boson
correlations, let us investigate how the correlation hole stipulated
by the repulsion between bosons at small separations changes due to
the influence of the surrounding bosons. At $n \to 0$ this hole is
completely specified by the condensate-condensate pair wave function
$\varphi(r)$. Exploring how ${U}(k)$ is expressed in terms of
${U}^{(0)}(k)$, makes it possible to know how $\varphi(r)$ differs
from $\varphi^{(0)}(r)$ at small boson separations. Note that the
relation connecting ${U} (k)$ with ${U}^{(0)}(k)$ has been published
in our previous paper~\cite{Ch1} without supporting calculations for
reasons of space.  Let us give these important calculations here.
Using the definition of ${U}(k)$ and Eq.~(\ref{12}), for the
scattering amplitude one can find
\begin{equation}
{U}(k)={\Phi}(k)-
      \frac{1}{2}\int\frac{\d^3q}{(2\pi)^3}\frac{{\Phi}(|{\bf k}
                  -{\bf q}|){U}(q)}{\sqrt{\widetilde{T}^2_q
                                +2n\widetilde{T}_q{U}(q)}}
\label{15}
\end{equation}
which can be called the in-medium Lippmann-Schwinger equation. Let us
rewrite Eq.~(\ref{15}) in the form $$
{U}(k)={\Phi}(k)-\frac{1}{2}\int\frac{\d^3q}{(2\pi)^3}
                 \frac{{\Phi}(|{\bf k}-{\bf q}|){U}(q)}{T_q}-I,
$$
where for $I$ we have
$$
I=\frac{1}{2}\int\frac{\d^3q}{(2\pi)^3}\Biggl(\frac{{\Phi}(|{\bf k}
  -{\bf q}|){U}(q)}{\sqrt{\widetilde{T}^2_q+ 2n\widetilde{T}_q{U}(q)}}-
  \frac{\Phi(|{\bf k}-{\bf q}|){U}(q)}{T_q}\Biggr).
$$
Performing the ``scaling" substitution~(\ref{subst})
in the integral and, then, taking the zero-density limit
in the integrand, for $n \to 0$ we find
\begin{equation}
I= -\alpha {\Phi}(k),
\quad \alpha=\frac{\sqrt{n m^3}}{\pi^2\hbar^3}{U}^{3/2}(0).
\label{16}
\end{equation}
From Eqs.~(\ref{15}) and (\ref{16}) it now follows that
\begin{eqnarray}
{U}(k)-{U}^{(0)}(k)=\alpha{\Phi}(k)
-\int&&\frac{\d^3q}{(2\pi)^3} \frac{{\Phi}(|{\bf k}-{\bf q}|)}{2T_q}
\nonumber \\
&&\times[{U}(q)-{U}^{(0)}(q)],
\label{17}
\end{eqnarray}
where $U^{(0)}(k)$ obeys Eq.~(\ref{15}) with $n=0$, i.e. the standard
Lippmann-Schwinger equation. Introducing the new quantity
${\xi}(q)=-[{U}(q) -{U}^{(0)}(q)]/(2T_q)$, for its Fourier transform
$\xi(r)$ we find the equation which is nothing else but the Schr\"odinger
equation (\ref{twobody}) with $\varphi^{(0)}(r)$ replaced by $\alpha+\xi(r)$.
As $\xi(r) \to 0$ when $r \to \infty$, then we can conclude that
$\xi(r)=\alpha \psi^{(0)}(r)$. Hence, for $n \to 0$ we get
\begin{equation}
{U}(k) \simeq {U}^{(0)}(k)\bigl(1 +\gamma(k,n)
\frac{8}{\sqrt{\pi}}\sqrt{na^3}\,\bigr).
\label{19}
\end{equation}
Here $\gamma(k,n) \to 1$ when $n\to 0$, and the scattering length $a$
is defined by Eq.~(\ref{adef}). The derived result for the in-medium
scattering amplitude $U(k)$ coincides with the low-density expansion
for the effective potential found within the effective-interaction
approach at the zero temperature (see Eq.~(4.27) in the review
\cite{shigrif}). This shows once more that there are actual parallels
between our model and approach of Ref.~\cite{Lee}. However, these
parallels are accompanied by significant differences. First, in
general the in-medium Lippmann-Schwinger equation (\ref{15}) is not a
variant of the t-matrix equation being frequency dependent contrary
to Eq.~(\ref{15}). Second, Eq.~(\ref{15}) has been found beyond any
diagram technique by means of the variational procedure whose
consequence is that the pair wave functions ``generating" in-medium
scattering amplitudes, coincide with the pair wave functions involved
in $g(r)$. On the contrary, it is not true for the
effective-interaction scheme. It implies the plane waves for
$\varphi_{{\bf p}}({\bf r})$ in the pair distribution function (see
the discussion in Sec.~\ref{effpot}) and certainly goes beyond the
plane-wave approximation when calculating t-matrix corresponding to a
pair of particles with a nonzero total momentum. Third, in
Eqs.~(\ref{12a})-(\ref{15}) we deal with $\widetilde{T}_k$ rather
than with $T_k$ appearing in the effective-interaction scheme. For
more information, see also discussion in the paper~\cite{Ch2}. Now,
returning to Eq.~(\ref{19}), we can conclude that for $r\lesssim R$
($R$ is the radius of the interaction potential $\Phi(r)$, for
strongly singular potentials $R$ is of the order of the scattering
length $a$) and $n\to 0$ we obtain the following in-medium
renormalization:
\begin{equation}
\varphi(r)\simeq
   \varphi^{(0)}(r)\Bigl(1+\frac{8\sqrt{na^3}}{\sqrt{\pi}}\Bigr).
\label{19a}
\end{equation}
Thus, the correlation hole coming from the repulsion of bosons at small
particle separations gets less marked with an increase of the density of the
surrounding bosons, which is mainly the result of the Bose-Einstein statistics. For
the pair distribution function at small boson densities we have $g(r) \propto
[\varphi^{(0)}(r)]^2$ [see the expression (\ref{grw})]. So, for
the strongly singular potentials, when $\varphi^{(0)}(r=0)=0,$ the correct
strong-coupling result $g(r=0)=0$ takes place for a dilute Bose gas if
Eqs.~(\ref{3}), (\ref{7}) and (\ref{rel}) are taken as the basic relations.

It is interesting to note that for the effective-interaction
approach~\cite{Lee} the upper cutoff in the momentum space at
$p_{c}\simeq\hbar/a$ is usually made. Thus, one could expect that
owing to the uncertainty relation the in-medium renormalization is
essential when $a\lesssim r$ in the real space. Equation (\ref{19a})
shows that this is not the case. As it is seen, $\varphi(r)$ at small
separations is really a solution of the ``bare" Schr\"odinger equation
(see Sec.~\ref{1sec}) but differs from
$\varphi^{(0)}(r)$ by the multiplier $(1+8\sqrt{na^3}/\sqrt{\pi})$.
We also remark that the peculiar overscreening takes place for
$\varphi(r)$ when $r\to \infty$. Indeed, Eq.~(\ref{12}) yields
$$
\lim_{k\to 0} k\psi(k)=-\frac{1}{2}\sqrt{\frac{m U(0)}{n}}.
$$
The last relation implies that for the condensate-condensate pair wave
function we have $\psi(r)=\varphi(r)-1\propto 1/r^{2}$ for $r\to
\infty$, in contrast to the ``bare" wave function
$\psi^{(0)}(r)=\varphi^{(0)}(r)-1\propto 1/r$ for $r\to \infty$
[see Eq.~(\ref{scatasymp})].

\section{Low-density expansions}
\label{5sec}

Now, to verify that a subtle balance of the terms coming from the
short-range particle correlations plays a significant role in the
problem of the strong-coupling Bose gas, let us calculate low-density
expansions of the basic thermodynamic quantities. The relation for
the condensate depletion
\begin{equation}
\zeta=\frac{n-n_0}{n}=\int \frac{\d^{3}q}{(2\pi)^{3}}\frac{n_{q}}{n}
=\frac{8\sqrt{na^3}}{3\sqrt{\pi}}+\cdots
\label{depletion}
\end{equation}
can be obtained from Eq.~(\ref{12a}) with the ``scaling" substitution
given by Eq.~(\ref{subst}).

The low-density expansion for the energy can be derived in four different
ways.

\subsection{The chemical potential}
\label{chempot}

The first way of obtaining the energy expansion deals with the
chemical potential $\mu$ and starts from the following relation for
$\mu$ valid in presence of the Bose condensate~\cite{Bog2}:
\begin{equation}
\mu=\frac{1}{\sqrt{n_0}}\int \d^3r'\,\Phi(|{\bf r}-{\bf r}'|)
\langle \hat{\psi}^{\dagger}({\bf r}')\hat{\psi}({\bf r}')
\hat{\psi}({\bf r})\rangle.
\label{24}
\end{equation}
Here $\hat{\psi}^{\dagger}({\bf r})$ and $\hat{\psi}({\bf r})$ stand
for the Bose field operators. This relation follows from the
well-known expression for an infinitesimal change of the grand
canonical potential
$
\delta\Omega=\langle\delta(\hat H-\mu\hat N)\rangle
$
and the necessary condition of the minimum of $\Omega$ with respect
to the order parameter $N_{0}$: $\partial\Omega(N_0,\mu,T)/\partial
N_0=0$, the Hamiltonian depending on the number of the condensed
particles owing to the substitution $\hat{a}_0^{\dagger}=\hat{a}_0=
\sqrt{N_0}$. Equations~(\ref{25}) and (\ref{24}) lead to~\cite{Not5}
\begin{equation}
\mu=n_0 {U}(0)
+\sqrt{2}\int\frac{\d^3q}{(2\pi)^3}n_q{U}_{{\bf q}/2}({\bf q}/2),
\label{26}
\end{equation}
where $U(0)$ is defined by Eq.~(\ref{11b}), and
\begin{equation}
{U}_{{\bf p}}({\bf k})=\int\d^3r\,\varphi_{\bf p}({\bf r})\Phi(r)
\exp(-i{\bf k}\cdot{\bf r}).
\label{26a}
\end{equation}
Using the substitution (\ref{subst}) in the integral and taking into
consideration Eqs.~(\ref{limphip}), (\ref{19}) and (\ref{depletion}),
we can rewrite Eq.~(\ref{26}) for $n \to 0$ as
\begin{equation}
\mu\!=\!n{U}(0)(1+\zeta+\cdots)=\frac{4\pi\hbar^2 an}{m}
\bigl(1+\frac{32}{3\sqrt{\pi}}\sqrt{na^3}+\cdots\bigr).
\label{26b}
\end{equation}
This, together with the thermodynamic relation $\mu=\partial
\bigl(n\varepsilon(n)\bigr)/\partial n$, yields the following result:
\begin{equation}
\varepsilon=\frac{2\pi\hbar^2 a n}{m}
\Bigl(1+\frac{128}{15\sqrt{\pi}}\sqrt{na^3}+\cdots\Bigr)
\label{23}
\end{equation}
known since the familiar paper by Lee and Yang~\cite{LY} and found
with the binary collision expansion method.

\subsection{Direct calculation of the energy}
\label{direct}

The way of this subsection is direct and starts from the
expression~(\ref{7}). Inserting Eq.~(\ref{3}) into Eq.~(\ref{8}) and
using the substitution (\ref{subst}) in the integral, we can rewrite
the pair distribution function for $n\to 0$ in the form
\begin{equation}
g(r)\simeq (1+2\zeta)\varphi^{2}(r),
\label{grlim}
\end{equation}
where the relation (\ref{limphip}) is implied. Note that this
expression is not valid at sufficiently large $r$ as the boundary
condition $g(r)\to 1$ for $r\to\infty$ is not satisfied. However,
here we are not interested in the long-range behaviour of $g(r)$
because we use Eq.~(\ref{grlim}) when integrating $g(r)$ multiplied
by the short-range potential $\Phi(r)$. Equation (\ref{grlim}) makes
it possible to represent Eq.~(\ref{7}) for $n \to 0$ as
\begin{eqnarray}
\varepsilon
&\simeq&\frac{n}{2}(1+2\zeta){U}(0)
\nonumber \\
&&+\int\frac{\d^3q}{(2\pi)^3}
\Bigl[T_q\frac{n_q}{n}+\frac{n}{2}(1+2\zeta) {U}(q){\psi}(q)\Bigr].
\label{21}
\end{eqnarray}
Taking the term proportional to $\zeta$ in the integral in
Eq.~(\ref{21}), we can rewrite it for $n \to 0$ in the form
\begin{eqnarray}
I=n\zeta\int\frac{\d^3q}{(2\pi)^3}U(q)\psi(q)
&\simeq&n\zeta\int\frac{\d^3q}{(2\pi)^3}{U}^{(0)}(q)\psi^{(0)}(q)
\nonumber \\
&=&-n\zeta\frac{4\pi\hbar^{2}}{m}b,
\label{intI}
\end{eqnarray}
where the vacuum scattering amplitude $U^{(0)}(q)$ and the
characteristic length $b$ are given by Eqs.~(\ref{U0}) and
(\ref{22b}), respectively. Using the substitution (\ref{subst}) in
the residual part of the integral in Eq.~(\ref{21}), $$
\int\frac{\d^3q}{(2\pi)^3}
\Bigl[T_q\frac{n_q}{n}+\frac{n}{2}U(q)\psi(q)\Bigr], $$ and taking
into account Eqs.~(\ref{12a}), (\ref{12}), (\ref{19}) and
(\ref{depletion}), we arrive at Eq.~(\ref{23}) but with the second
term multiplied by the factor $\lambda=1-5b/(8a)$. Now the question
arises which variant we should prefer, Eq.~(\ref{23}) or
Eq.~(\ref{23}) with the factor $\lambda=1-5b/(8a)$, and what is the
reason for this ambiguous situation?

To answer this question, let us reconsider the procedure of
calculating $\varepsilon$ given in this subsection. As it has been
mentioned earlier, Eqs.~(\ref{12a}) and (\ref{12}) used in our
calculations are consistent with Eq.~(\ref{bogrel}) rather than with
Eq.~(\ref{rel}). Being characteristic of the Bogoliubov model,
Eq.~(\ref{bogrel}) is accurate to the leading order in $(n-n_0)/n$
and differs from Eq.~(\ref{rel}) by the supracondensate-condensate
term neglected in the Bogoliubov relation. The problem of the
$\lambda$-factor turned out to be directly related to this term. It
can be taken into account by representing Eq.~(\ref{rel}) for $n \to
0$ in the form
\begin{equation}
n_{k}(n_{k}+1)=(1+2\zeta)\psi^{2}(k).
\label{connect1}
\end{equation}
Solving this equation with respect to $n_{k}$ and noticing
Eq.~(\ref{12}), one can obtain
\begin{equation}
n_k=\frac{1}{2}
\Biggl(\frac{\sqrt{\bigl(\widetilde{T}_k+nU(k)\bigr)^{2}
                   +2\zeta n^{2}U^{2}(k)}}
            {\sqrt{\widetilde{T}_k^2+2n\widetilde{T}_k U(k)}}
       -1\Biggr).
\label{nk1}
\end{equation}
Now, restarting from Eq.~(\ref{21}) and making use of the system of
Eqs.~(\ref{nk1}) and (\ref{12}) instead of that of Eqs.~(\ref{12a})
and (\ref{12}), we arrive at Eq.~(\ref{23}). The term given by
(\ref{intI}) is now cancelled due to the correcting term $2\zeta
n^{2}U^{2}(k)$ involved in Eq.~(\ref{nk1}). So, we face a rather
complicated situation: namely, to get the correct result (\ref{23}) in
the direct calculations starting from Eq.~(\ref{7}) , we have to
abandon Eq.~(\ref{12a}) in favour of Eq.~(\ref{nk1}) while for
$\psi(k)$ we can exploit Eq.~(\ref{12}). The most important point
here is the uniform convergence of the integral in Eq.~(\ref{21})
provided Eq.~(\ref{nk1}) is used. This allows for employing
Eq.~(\ref{nk1}) together with Eq.~(\ref{12}) in spite of the fact
that the latter has been found in the leading order in $(n-n_0)/n$.
The higher-order corrections to Eq.~(\ref{12}) does not influence the
result of calculating the integral in Eq.~(\ref{21}) if we limit
ourselves to the leading and next-to-leading orders in $n a^3$. It is
worth noting that replacing Eq.~(\ref{nk1}) by Eq.~(\ref{12a}) does
not influence on Eqs.~(\ref{depletion}) and (\ref{26b}). So, the
preliminary result for $\varepsilon$ found in the paper~\cite{Ch1}
and corresponding to Eq.~(\ref{23}) with the second term multiplied
by the factor $\lambda=1-5b/(8a)$, has to be abandoned in favour of
Eq.~(\ref{23}).

The analysis fulfilled in this section demonstrates the crucial role
of the subtle balance of the terms coming from the boson scattering (or,
in other words, from the short-range boson correlations). Disturbance
of this fine interplay which seems to be insignificant, can
nevertheless lead to wrong conclusions. We stress that the
strong-coupling model of Ref.~\cite{Ch1} is balanced because it takes
into consideration the supracondensate-condensate scattering waves in
{\it both} the pair distribution function and relation connecting the
momentum distribution with the pair wave functions. On the contrary,
the effective-interaction approach is not balanced with respect to
the supracondensate-condensate scattering waves which are missed in
the pair distribution function but make a contribution to the
``dressed" potential~(see Sec. \ref{effpot}). Exactly this problem is
the basis for the so-called ultraviolet divergence occurring in the
effective-interaction approach.

\subsection{The energy expansion through the Hellmann-Feynman theorem}
\label{hellfeyn}

As it is shown in Sec.~\ref{effpot}, the effective-interaction
approach results in the irrelevant picture of the short-range boson
correlations. This is why it can not yield the correct individual
values of the interaction $\varepsilon_{int}$ and kinetic
$\varepsilon_{kin}$ energies by the direct calculations based on
Eq.~(\ref{7}). Recall that we have the following definitions:
these energies are defined by
\begin{eqnarray}
&&\varepsilon_{int}\!=\!\frac{1}{N}
\Bigl\langle\sum_{i\not=j}\frac{\gamma}{2}
\Phi(|{\bf r}_{i}-{\bf r}_{j}|)\Bigr\rangle
\!=\!\frac{n}{2}\int\d^{3}r\,\gamma\Phi(r)g(r),
\label{intdef}\\
&&\varepsilon_{kin}\!=\!\frac{1}{N}
\Bigl\langle-\sum_{i}\frac{\hbar^{2}\nabla_{i}^{2}}{2m}\Bigr\rangle
=\int\frac{\d^{3}k}{(2\pi^{3})}T_{k}\frac{n_{k}}{n},
\label{kindef}
\end{eqnarray}
where $\langle\cdots\rangle$ stands for the statistical average with
respect to the ground state, and the auxiliary parameter $\gamma$ is
the coupling constant. The total energy per particle (\ref{7}) is
given by the sum of $\varepsilon_{int}$ and $\varepsilon_{kin}$ at
$\gamma=1$:
\begin{equation}
E/N=\varepsilon=\varepsilon_{kin}+\varepsilon_{int}.
\label{entotal}
\end{equation}

Our model provides the correct short-range behaviour of the pair distribution
function $g(r)$. So, we can first evaluate $\varepsilon_{int}(\gamma)$, and,
then, obtain the total energy (\ref{7}) by means of the well-known expression
often called the Hellmann-Feynman theorem which is just the variational
theorem for the ground state obeying the $N$-body Schr\"odinger equation:
\begin{equation}
\delta E=\langle\delta\hat{H}\rangle. \label{29}
\end{equation}
In Eq.~(\ref{29}) $\delta E$ and $\delta\hat{H}$ are infinitesimal
changes of the average energy ($E=\langle\hat{H}\rangle$) and the
Hamiltonian
\begin{equation}
\hat{H}=-\sum_{i}\frac{\hbar^{2}\nabla_{i}^{2}}{2m}
+\frac{1}{2}\sum_{i\not=j}\gamma\Phi(|{\bf r}_{i}-{\bf r}_{j}|),
\label{ham}
\end{equation}
respectively. The relations (\ref{intdef}), (\ref{kindef}),
(\ref{29}) and (\ref{ham}) leads to the important equations:
\begin{eqnarray}
\varepsilon_{int}=\gamma\frac{\partial\varepsilon}{\partial\gamma},
\quad\varepsilon_{kin}=-m\frac{\partial\varepsilon}{\partial m}.
\label{intkin}
\end{eqnarray}
From the first expression in Eq.~(\ref{intkin}) it follows that
\begin{equation}
\varepsilon=\int_{0}^{1}\d\gamma\frac{\varepsilon_{int}(\gamma)}{\gamma}.
\label{energyHF}
\end{equation}
To evaluate $\varepsilon_{int}(\gamma)$ in the leading and
next-to-leading orders in $n a^3$, it is convenient to rewrite
Eq.~(\ref{grlim}) as
\begin{equation}
g(r)\simeq w(na^{3})[\varphi^{(0)}(r)]^{2}, \label{grw}
\end{equation}
where Eqs.~(\ref{19a}) and (\ref{depletion}) are taken into account
and $w(n a^3)$ is given by
\begin{equation}
w(na^{3})=1+\frac{64}{3\sqrt{\pi}}\sqrt{na^3}.
\label{wna3}
\end{equation}
The range of particle separations for which Eq.~(\ref{grw}) is
correct, coincides with that of Eq.~(\ref{grlim}) [see the
discussion there]. Keeping in mind Eqs.~(\ref{31int}), (\ref{33}),
(\ref{intdef}), (\ref{energyHF}) and (\ref{grw}), we obtain
$$
\varepsilon\!\simeq\! \frac{2\pi\hbar^{2}n}{m}\int_{0}^{1}\d\gamma
\frac{\partial a}{\partial \gamma}w(na^{3}(\gamma))
=\frac{2\pi\hbar^{2}n}{m}\int_{0}^{a}\d a'\, w(n{a'}^{3}).
$$
Thus, using Eq.~(\ref{wna3}), we arrive at Eq.~(\ref{23}) again.

\subsection{The energy expansion through the virial theorem}
\label{virial}

This method was proposed by Bogoliubov in his original
paper~\cite{Bog1} in order to obtain the leading-order term in the
energy expansion.  Here we consider this method in a more general form.
As in Sec.~\ref{hellfeyn}, we start from the expression for the pair
distribution function $g(r,n)$ which is assumed to be a known function of the
density $n$. The basic idea is to derive the differential equation
for $\varepsilon(n)$.

On the one hand, from the virial theorem we get the following
expression for the pressure
\begin{equation}
P=\frac{2}{3}\varepsilon_{kin}(n)n -\frac{n^{2}}{6}
\int\d^{3}r\frac{\d\Phi(r)}{\d r}r g(r,n),
\label{virialth}
\end{equation}
where $\varepsilon_{kin}(n)$ is given by Eq.~(\ref{kindef}). On the
other hand, we have the thermodynamic relation
\begin{equation}
P=n^{2}\frac{\partial \varepsilon(n)}{\partial n} \label{pressure}
\end{equation}
valid at the zero temperature. Here $\varepsilon(n)$ is the energy
per particle (\ref{7}) which can be written as
\begin{equation}
\varepsilon(n)=\varepsilon_{kin}(n)
                      +\frac{n}{2}\int\d^3r\,g(r,n)\Phi(r).
\label{energyn}
\end{equation}
The system of Eqs.~(\ref{virialth})-(\ref{energyn}) yields the
differential equation for $\varepsilon(n)$ whose general solution is
of the form:
\begin{equation}
\varepsilon=C_{0}n^{2/3}-
\frac{1}{6}\int\d^{3}r
\Bigl[\frac{\d\Phi(r)}{\d r}r + 2\Phi(r)\Bigr]\chi(r,n),
\label{energychi}
\end{equation}
where the function $\chi(r,n)$ stands for
\begin{equation}
\chi(r,n)=n^{2/3}\int_{0}^{n}\d n'\frac{g(r,n')}{{n'}^{2/3}}
\label{chi}
\end{equation}
and $C_{0}$ is the integration constant. Note that Eqs.~(\ref{energychi}) and
(\ref{chi}) are valid for both the Bose and Fermi systems because we have not
used the type of the statistics when deriving these equations. For a Fermi
system the constant $C_{0}$ is not equal to zero while for a Bose one we
should put $C_{0}=0$ provided the Bose-Einstein condensation takes place.
Substituting Eq.~(\ref{grw}) in Eq.~(\ref{chi}), from Eq.~(\ref{energychi})
we get
\begin{equation}
\varepsilon=J\frac{n^{2/3}}{6}\int_{0}^{n}\d
                          n'\frac{w(n'a^{3})}{{n'}^{2/3}},
\label{energy1}
\end{equation}
where $$ J=-\int\d^{3}r\Bigl[\frac{\d\Phi(r)}{\d r}r + 2\Phi(r)\Bigr]
[\varphi^{(0)}(r)]^{2}=U^{(0)}(0). $$ The last equation can be
derived after small algebra with the help of
Eqs.~(\ref{twobody})-(\ref{u0bare}). As it is
seen, Eq.~(\ref{energy1}), together with Eq.~(\ref{wna3})
leads to the low-density expansion (\ref{23}).

Thus, all the four ways of calculating $\varepsilon$ within the
strong-coupling model developed in Ref.~\cite{Ch1} leads to
Eq.~(\ref{23}).

\subsection{The interaction and kinetic energies}
\label{intkinsec}

For any physical quantity there usually exist various calculating
procedures leading to the same result provided the model considered
is consistent. By contrast, in presence of a thermodynamic
inconsistency different ways of calculating any thermodynamic quantity
are able to produce different results~\cite{Sh1,Sh2}, during which it
is often happened that one of them is reasonable but others are
completely inadequate. Such a situation is realized when evaluating
the interaction and kinetic energies via the effective-interaction
method. This is demonstrated below.

The interaction (\ref{intdef}) and kinetic (\ref{kindef}) energies of
a dilute Bose gas can be evaluated on the basis of the
Hellmann-Feynman theorem with the help of Eq.~(\ref{23}).
Representing this expansion for $\varepsilon$ in the form
\begin{equation}
\varepsilon=\frac{2\pi\hbar^2an}{m}f(na^3)\;, \label{35}
\end{equation}
and keeping in mind Eqs.~(\ref{intkin}) and (\ref{33}), one can
derive
\begin{eqnarray}
&&\varepsilon_{int}=\frac{2\pi \hbar^2 (a-b) n}{m} \left(
f(na^3)+3na^3 \frac{\d f(na^3)}{\d(na^3)}\right),
\label{36int} \\
&&\varepsilon_{kin}=\frac{2\pi\hbar^2b n}{m}\!\left(f(na^3)+
3na^3\!\left(1-\frac{a}{b}\right)\frac{\d f(na^3)}{\d(na^3)}\right).
\label{36}
\end{eqnarray}
According to Eq.~(\ref{23}) $f(x)=1+128\sqrt{x}/(15\sqrt{\pi})$,
which together with Eqs.~(\ref{36int}) and (\ref{36}) yields
\begin{eqnarray}
&&\varepsilon_{int}=\frac{2\pi \hbar^2 (a-b)n}{m}
\left(1+\frac{64}{3\sqrt{\pi}} \sqrt{n a^3}+ \cdots \right),
\label{34int}\\
&&\varepsilon_{kin}=\frac{2\pi \hbar^2 b n}{m}
\left(1+\frac{64}{3\sqrt{\pi}}\sqrt{n a^3}
\left(1-\frac{3a}{5b}\right)+ \cdots \right).
\label{34}
\end{eqnarray}
As it is seen from Eqs.~(\ref{35})-(\ref{36}), the terms involving
$b$ are present in the expressions for the kinetic and interaction
energy and mutually cancelled in the total energy $\varepsilon$.
We emphasize that the reasoning of this paragraph can be fulfilled
for both the effective-interaction approach and model developed by
the present authors.

Our approach is fully consistent, which makes it possible to derive
Eqs.~(\ref{34int}) and (\ref{34}) in another way using the direct
calculations. Indeed, Eq.~(\ref{intdef}) taken at $\gamma=1$ in
conjunction with Eqs.~(\ref{31int}) and (\ref{grw}) results in
Eq.~(\ref{34int}). Notice that the supracondensate-condensate
scattering waves make a significant contribution to the
next-to-leading term of the low-density expansion
for $\varepsilon_{int}$. It is not also difficult to find the
low-density expansion for the kinetic energy (\ref{34}) with the
help of Eqs.~(\ref{19}), (\ref{depletion}) and
(\ref{nk1}).

On the contrary, due to the thermodynamic inconsistency the
effective-interaction scheme does not allow for obtaining
Eqs.~(\ref{34int}) and (\ref{34}) directly, beyond the
Hellmann-Feynman theorem taken with Eq.~(\ref{23}).
Let us show this for the pseudopotential
approach discussed in Sec.~\ref{effpot}. To evaluate the kinetic
energy in this case, one can start from Eq.~(\ref{kindef}) and use
the Bogoliubov formula (\ref{21aa}) with the pseudopotential
substitution (\ref{hardsp}). In so doing the divergent integral
$\int\d^{3}k/k^{2}$ should be ignored [see the discussion in
Sec.~\ref{effpot}]. Similarly, the interaction energy at $\gamma=1$
can be derived from Eqs.~(\ref{11}), (\ref{21aa}), (\ref{psibog}) and
(\ref{hardsp}) using the same regularization. However, to simplify
the calculations, we adopt the other way leading to the same results
and based on the low-density expansion (\ref{21i}) found within the
Bogoliubov model. The original Bogoliubov scheme is fully consistent,
which implies equivalence of different ways of calculating any
thermodynamic quantity. Therefore, we can first find the kinetic and
interaction energy by using the Hellmann-Feynman theorem together
with Eq.~(\ref{21i}) and, then, replace $a_0$ by $a$ and $a_1$ by $0$
in the derived expressions. From the definition (\ref{borna}) it
follows that
$$ \gamma\frac{\partial a_{0}}{\partial \gamma}=
m\frac{\partial a_{0}}{\partial m}=a_{0},\quad \gamma\frac{\partial
a_{1}}{\partial \gamma}= m\frac{\partial a_{1}}{\partial m}=2a_{1}.
$$
Hence, within the Bogoliubov model we can arrive at
\begin{eqnarray}
&&\varepsilon_{int}=\frac{2\pi\hbar^2n}{m}
\left(a_{0}+2a_{1}+a_{0}\frac{64}{3\sqrt{\pi}}\sqrt{n a_{0}^{3}}+
\cdots\right),
\label{bogint}\\
&&\varepsilon_{kin}=\frac{2\pi\hbar^2 n}{m}
\left(-a_{1}-a_{0}\frac{64}{5\sqrt{\pi}}\sqrt{n a_{0}^3}
+ \cdots \right),
\label{bogkin}
\end{eqnarray}
provided Eqs.~(\ref{21i}) and Eq.~(\ref{intkin}) are taken into
consideration. Now, replacing $a_{0}$ by $a$ and substituting $0$
for $a_{1}$~(the latter allows for escaping the ultraviolet
divergence, see Sec.~\ref{effpot}), we obtain the following expressions:
\begin{eqnarray}
&&\varepsilon_{int}=\frac{2\pi\hbar^2 a n}{m}
\left(1+\frac{64}{3\sqrt{\pi}}\sqrt{n a^{3}}+
\cdots\right),
\label{hspint}\\
&&\varepsilon_{kin}=-\frac{2\pi\hbar^2 a n}{m}
\frac{64}{5\sqrt{\pi}}\sqrt{n a^3}+ \cdots
\label{hspkin}
\end{eqnarray}
which should be compared with the correct results given by
Eqs.~(\ref{34int}) and (\ref{34}). As it is seen, the sum of the
r.h.s. of Eqs.~(\ref{hspint}) and (\ref{hspkin}) gives the r.h.s. of
Eq.~(\ref{23}) but at the expense of the negative value of the
kinetic energy (\ref{hspkin}). Notice that the Bogoliubov model is
free from this nonphysical feature because $a_{1}<0$, which leads to
$\varepsilon_{kin}>0$ at sufficiently small densities. Thus,
Eqs.~(\ref{hspint}) and (\ref{hspkin}) found within the
pseudopotential model are inadequate.  The reason is obvious: the
pseudopotential scheme allows for restoring the functional dependence
on the scattering length $a$ in Eqs.~(\ref{hspint}) and
(\ref{hspkin}) while completely ignores the additional length $b$
that cannot be involved in the pseudopotential model due to the
ultraviolet divergence.

Note that in the case of the hard-sphere interaction  (\ref{hardsph})
we get $a=b$ from the solution of the Schr\"odinger equation
(\ref{twobody}). Then, in the general case of Eq.~(\ref{35}) the
relations $\varepsilon_{int}=0$, $\varepsilon_{kin}=\varepsilon$ come
from Eqs.~(\ref{36int}) and (\ref{36}). Thus, we arrive at the
interesting property of the Bose gas with the pairwise potential
(\ref{hardsph}): namely, although the Bose gas is strongly
interacting, the interaction energy is equal to zero. Hence, the
total energy of a dilute Bose gas made of the hard spheres is exactly
equal to the kinetic one. One can see that this result is rather
general: {\it for the hard-sphere potential (\ref{hardsph}) the
interaction energy is equal to zero for any density.} Indeed,
$\Phi(r)$ given by (\ref{hardsph}) can be thought of as the limiting
case of the repulsive potential
\[
\Phi(r)=\left\{\begin{array}{ll}
V_{0},   & r<a, \\
0,       & r>a.
\end{array}\right.
\]
It is clear that the saturation takes place when $V_{0}\gg
\varepsilon$: further increase of the parameter $V_{0}$ does not
change the energy per particle $\varepsilon$. Hence, according to
Eqs.~(\ref{intdef}) and (\ref{intkin}), $\varepsilon_{int}=0$ because
$\partial\varepsilon/\partial \gamma=0$ at $\gamma=1$ in the limit $V_{0}\to
+\infty$. Notice that even taken in the order linear in the density $n$,
Eqs.~(\ref{hspint}) and (\ref{hspkin}) lead to the opposite case
$\varepsilon_{int}\simeq \varepsilon$, $\varepsilon_{kin}\simeq 0$.  This
incorrect redistribution of the energy of a dilute Bose gas in the
pseudopotential approach is also remarked in the paper~\cite{Lieb} where the
leading order of the low-density energy expansion is considered.

Note that the relation Eq.~(\ref{33}) enables us to obtain the length
$b$ in an experimental way, by the isotopic shift of the scattering
length $a$:
\[ b=a\Bigl(1-\frac{\partial \ln a}{\partial \ln m}\Bigr)\simeq
a\Bigl(1-\frac{\Delta a}{a}\frac{m}{\Delta m}\Bigr).
\]
Hence, we are able to evaluate the interaction (\ref{34int}) and
kinetic (\ref{34}) energies per particle via the quantities which
can be found experimentally.
\section{Conclusion}
\label{7sec}

In conclusion, we remark that this paper concerns the thermodynamics
of a dilute Bose gas with a strongly repulsive interaction in the leading
and next-to-leading orders of the low-density expansion. The
strong-coupling generalization of the Bogoliubov model proposed by
the present authors is shown to reproduce the result (\ref{23}) of
Lee and Yang~\cite{LY} found via the binary collision expansion
method. Contrary to the effective-interaction approach of
Ref.~\cite{Lee}, the model considered in this paper is
thermodynamically consistent and free of the ultraviolet divergences.
These advantages are due to accurate treatment of the short-range
spatial boson correlations whose picture is inadequate within the
effective-interaction scheme. The present paper thus demonstrates
that the effective-interaction scheme which is reduced to the
Bogoliubov model with the effective pairwise potential, is not
acceptable for investigating a dilute strongly interacting Bose gas.
In addition to the arguments mentioned above, this also follows from
the results for the kinetic and interaction energies first found in
this paper.

In some sense the strong-coupling model discussed can be considered
as a generalization of the Brueckner approach taken in its
representation given by Bethe and Goldstone~\cite{Bethe}. The new
essential point is that the in-medium pair wave functions are
calculated {\it in conjunction} with the particle momentum
distribution on the basis of the variational procedure. So, to go
further, additional investigations should be fulfilled to establish
more accurate approximations of the relation connecting the boson
momentum distribution with the scattering parts of the in-medium pair
wave functions. In particular, this improvement is needed to clarify
to what extent the correct spectrum of elementary excitations in a
dilute Bose gas differs from the well-known prediction of the
effective-interaction approach. Of course, the region of intermediate
momenta is implied rather than the linear phonon sector which should
be the same according to the thermodynamic prescription. This problem
is tightly related to investigation of the long-range spatial boson
correlations beyond the effective-interaction approach.

This work was supported by the RFBR grant No. 00-02-17181.


\end{multicols}

\begin{references}
\bibitem{Not3a} It is not, of course, the case at nonzero
temperatures when spatial boson correlations take place even in
absence of any interaction. Nevertheless, this difference between the
zero- and nonzero-temperature situations does not influence the fact
that $n_k$ is connected with the scattering parts of the pair wave
functions (\ref{4}).
\bibitem{Not0} By the leading order in $(n-n_0)/n$ we mean the zero
depletion of the condensate: $n=n_0$.
\bibitem{Not2} Using formula (5.12) of the paper of Hugenholtz
and Pines~\cite{Lee} together with the relations (4.4) of the paper
of Beliaev~\cite{Lee}, for $n \to 0$ one can derive $S(k)\to
1-n{U}^{(0)}(k)/T_k$, where ${U}^{(0 )}(k)$ is
the Fourier transform of $U^{(0)}(r)= \varphi^{(0)}(r) \Phi(r)$ and
$T_k=\hbar^2 k^2/2m$. So, using the definition of the structural
factor, one can arrive at $g(r) \to 1+2{ \psi}^{(0)}(k)$
for $n \to 0$. Here for ${\psi}^{(0)} (k)$ we have $-2T_k
{\psi}^{(0)}(k)={U}^{(0)} (k)$, which can be
rewritten in the form of Eq.~(\ref{twobody}).
\bibitem{Not4} We stress that the potential is involved through the
mediation of the scattering length (\ref{adef}) only in a few first
terms in the low-density expansions. For the high-order terms the
dependence on a specific shape of the potential should appear.
\bibitem{Note5a} The Bogoliubov model which is in fact the perturbation
theory with respect to the coupling constant $\gamma$, provides the
accuracy up to the last term in Eq.~(\ref{21i}) proportional to
$\gamma^{5/2}$~\cite{Tolmach}. For this reason, in the expansion
(\ref{21i}) there is no $(\partial c_{2}/\partial a) a_{1}$ term
proportional to $\gamma^{7/2}$.
\bibitem{Not5} Note that Eq.~(\ref{26}) can help to derive the
strong-coupling generalization of the Gross-Pitaevskii equation on the solid
theoretical basis beyond any effective-interaction picture.
\end{references}
\end{document}